# A Trajectory UML profile For Modeling Trajectory Data: A Mobile Hospital Use Case

Wided Oueslati[1], Jalel Akaichi[2]
University of Tunis
Institut Supérieur de Gestion
41, Rue de la Liberté, Cité Bouchoucha
Le Bardo 2000
Tunisia
[1]widedoueslati@live.fr, [2]jalel.akaichi@isg.rnu.tn

*Abstract*: A large amount of data resulting from trajectories of moving objects activities are collected thanks to localization based services and some associated automated processes. Trajectories data can be used either for transactional and analysis purposes in various domains (heath care, commerce, environment, etc.). For this reason, modeling trajectory data at the conceptual level is an important stair leading to global vision and successful implementations. However, current modeling tools fail to fulfill specific moving objects activities requirements. In this paper, we propose a new profile based on UML in order to enhance the conceptual modeling of trajectory data related to mobile objects by new stereotypes and icons. As illustration, we present a mobile hospital use case.

## I. INTRODUCTION

In many countries, rural populations undergo various inadequacies related to health care necessities. In fact, due to health care problems, they have to cross long distances to reach medical structures such as hospitals. This, obviously, may generate loss of lives and/or medical complications. Mobile hospitals, which may contribute to decrease such misery, can move to populations, following declared emergencies and/or planned activities.

The notion of mobile hospitals was emerged then to try to decrease the geographical disparities related to health care services. It takes profits of the emergence of adapted medical hardware, positioning technologies and pervasive systems. In fact, technologies of mobile communications and ubiquitous computing are saturating our new world. Wireless networks are becoming the nerves of our territory; through these nerves, the movement of people and vehicles may be sensed and possibly recorded, thus producing large volumes of mobility data.

Modeling the trajectory of the mobile object, such as a mobile hospital in our case, represents real challenge. Indeed, witnessing the trajectory of a moving hospital implies the recording of different information leading to get a finite set of observations destined to current transactional and future analysis activities.

The mobility of a hospital, supported by adequate terrestrial transportation mean, is constrained by the road network. Mobile hospitals manager defines for each mobile hospital its trajectory which is composed by a set of stops and moves. For each trajectory, the mobile hospital is equipped with medical hardware, physicians, nurses, drivers, and a local mobile hospital manager. This latter is equipped with a PDA used to manage trajectory data. He has to indicate the beginning and the end of the trajectory, each stop and move. He also sends to or receives from messages to the responsible of the mission. Physicians, equipped equally with PDA, sends data related to patients and examinations to a centralized data warehouse. Such data are important since they allow detecting prospective epidemic peaks.

Modeling trajectory data at the conceptual level is very important since it leads to a global vision and successful implementations. However, currents modeling tools fail to fulfill specific moving objects, such as mobile hospital, activities requirements. In this paper, we propose a new profile based on UML in order to enhance the conceptual modeling of trajectory data related to mobile objects by new stereotypes and icons. As illustration, we present a mobile hospital use case.

This paper is organized as follows: In section 2, we will present different research works related to moving objects (definition of moving objects and different types of those latter), trajectory data conceptual model and positioning technologies. In section 3, we will propose a trajectory data conceptual modeling for mobile hospitals. In section 4, we will present the implementation of our trajectory data profile based on UML. In section 5, we will summarize the work and give some perspectives to be done in the future.

## II. STATE OF THE ART

A moving object is considered as a spatial object which changes its location over time. It is characterized by its numerous positions, which vary in space and time, describing a trajectory. The trajectory of a moving object is considered then as a path since it reflects the evolution of an object travelling in some space during a given time interval.

Some works [1, 2, and 3] define a trajectory as a set of stops and moves. Each element of a set has a "begin" and an "end". The "stop" is considered as an important component of the trajectory which is characterized by a non-empty time interval. This means that the moving object can perform many actions but it has to be immovable in a given "stop". A move is delimited by two stops between which a moving object is in movement. Authors in [4] distinguish many types of trajectories according to relationships between them or with their environments. For the first type of relationship, we can cite those relations: Intersect (two trajectories cross themselves), Equal (a trajectory is near another trajectory), Near (a trajectory is near another trajectory) and Far (a trajectory is far from another trajectory).

For the second type of relationship, trajectories can be in relation with other spatial objects such as infrastructure elements (roads, buildings, etc.) or virtual entities such as border of a city. Among those relationships we can cite: Stay within (the trajectory is always in a sector of interest), Bypass (the trajectory pass by the sector of interest), Leave (the trajectory leaves the sector of interest), and Enter (the trajectory enters the sector of interest) and Cross (the trajectory crosses the sector of interest).

To collect data resulting from moving objects activities, positioning and sensors technologies, and services associated to them and technologies with sensors seem to be a necessity. In fact, they permit to gather, to process and to broadcast localization information and data generated in it. Among those technologies we can mention GPS (Global Positioning System), RFID (Radio Frequency IDentification tags), Bluetooth, WiFi, ZigBee, etc., which permit the localization and the follow up of persons, goods and animals on move.

Positioning technologies can be classified into two categories according to their mobility degree [5]: fixed positioning technologies and mobile ones.

Positioning technologies [5, 6, 7] can have many functions such as authentication, data transfer via smart cards for example, object localization, etc. Services associated to them are responsible of the acquisition of data for a given system which is able to collect data coming from his environment.

Many research works were interested in the conceptual modeling of data and data warehouses [8, 9, and 10], spatial and temporal data and data warehouses [11, 12, and 13] but few of works were interested in the conceptual modeling of trajectory data and trajectory data warehouses [1, 14]. In [1], the requirements for the trajectory modeling take into account the characterization of trajectories and their components and the different types of constraints (semantic, topologic, etc.) in order to fix a conceptual view of the concept of trajectory. The conceptual model is seen as a direct support and an explicit representation of trajectories' components (stops, moves, begins, ends).

The two solutions proposed in [1] are driven by different modeling goals. The first solution suggests a trajectory design pattern, and the second one recommends some trajectory data types. The two approaches can be combined to offer richer modeling tools when needed. The design pattern is a predefined generic schema that can be connected to any other database schema by the designer and can be adjustable. In fact, the designer can modify the design pattern by adding new elements or semantic attributes, and/or deleting other ones to adapt it to the requirements of the new applications. The second approach is based on trajectory data types. It holds the idea that semantic information is specific to an application and cannot be encapsulated into a data type, but has to be defined by the database designer.

Moreover, authors in [1] define a generic data types to hold the trajectories' components such as the "begin", the "end", the "stops" and the "moves", etc., and functions of interpolation. We have to mention that authors in [1], extended those data models for trajectories from MADS model [15], since this latter supports spatial and temporal objects and relationships. Indeed, they announce icons defined in it, in their trajectory data relational conceptual model to make it more understandable by users.

## III. TRAJECTORY DATA CONCEPTUAL MODELING

### A. Mobile hospital's states conceptual modeling

The trajectory of the mobile hospital is composed by a set of "moves" and "stops". The mobile hospital can be in the states "ready", "move in road", "stop failure", "stop in destination" and "move in destination". It is ready when everything and everyone are in the transportation platform and the driver is ready to start his mission (to move in his trajectory). It is in "move in road" when everything and everyone are in the transportation platform which is in motion in the road network. It is "stop failure" when the transportation platform is stopped because of broke down or any other problem. It is in "stop in destination" when the mobile hospital arrives to the targeted destination which is a place belonging to mobile hospital trajectory fixed by the responsible of the mission. It is in "move in destination" when the mobile hospital moves around the destination to find some emergencies in cases of critical state of patients or to find the nearest health care institution encapsulating sophisticated materials and/or care givers specialists that don't exist in the mobile hospital. We propose the following schema to describe the different states of the mobile hospital during his trajectory.

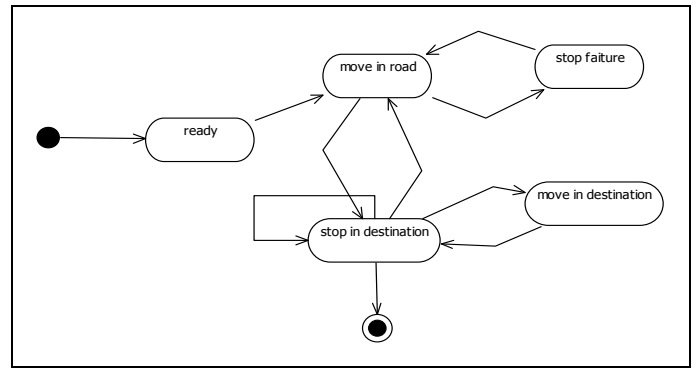

Figure 1. Different states of the mobile hospital.

### B. Medical staff's tasks conceptual modeling

In this section, we present some tasks which are done by the medical staff such as doctors. Those latter can consult the table "patient" or add a new patient or add new description in the table "patient". In fact, the doctor can simply wrote the patient name that he wants to consult, the system checks the presence of this patient and display the patient information. Finally the system display patient information. After the consultation, the doctor can add a prescription for example to the table "patient". In fact, the doctor wrote the patient prescription, and then the system sends the added information to the table "patient". We propose the following sequence diagrams that illustrate the description cited above:

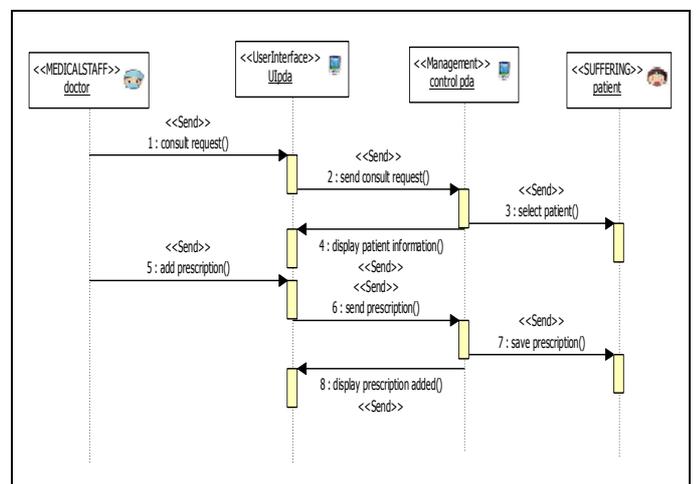

Figure 2. Consult and modify patient sequence diagram .

## C. *Mobile hospital trajectory data conceptual modeling*

The conceptual modeling of trajectory data resulting from a mobile hospital on road is represented by a UML class diagram. In the following, we describe classes which model the trajectory data and the different relationships between them:

- A mobile hospital class is composed by medical machines and staff sub_classes. Each mobile hospital has its own trajectory which is composed by a set of moves and stops.
- A Trajectory class is considered as a travel in the space and in the time. It is composed by a set of trajectory sections. It is characterized by a set of attributes such as the "begin", the "end" and the duration of a given trajectory.
- A Trajectory-section class is a component of the whole trajectory. Each trajectory section is composed by two successive "stops" and one "move".
- A Stop class is modeled by its position in the space and its time of "begin" and "end". The "begin" of a "stop" is the "end" of a previous "move". As the "stop" is the core component of the trajectory, we propose the following sequence diagram to show how to specify a given stop.
- A Move class is delimited by two successive "stops". The move is modeled by the position of the "begin" and of the "end" of a movement. Those positions are represented by absolute coordinates which are variable in time. As the move is the core component of the trajectory, we propose the following sequence diagram to show how to specify a given move in the system.
- A Patient class describes the person treated in the mobile hospital by medical staff.
- A Doctor class describes the person who is working in the mobile hospital.
- A Nurse class describes the person who is working in the mobile hospital.
- A Driver class describes the person who is driving the mobile hospital.
- A GPS-data class can be connected to a given PDA and has an identifier and a set of attributes like the latitude and the longitude
- A PDA class is the mean of sending trajectory data. In fact, each manager and doctor has one PDA to send trajectory data and to communicate with the responsible of his mission. The PDA has an identifier and it is connected to a GPS.
- A mobile hospital manager class describes the person who is sending trajectory data using a given PDA.

The following class diagram is related to trajectory data resulting from mobile hospital. We created a new UML profile called Trajectory-UML, in which we added some pictograms and some stereotypes to identify each class (entity). We used pictograms of MADS project [15] only for some entities which are representing trajectory and trajectory components such as "move", "stop", "tr-section", and "location". The same idea was done in [1] but in a relational model.

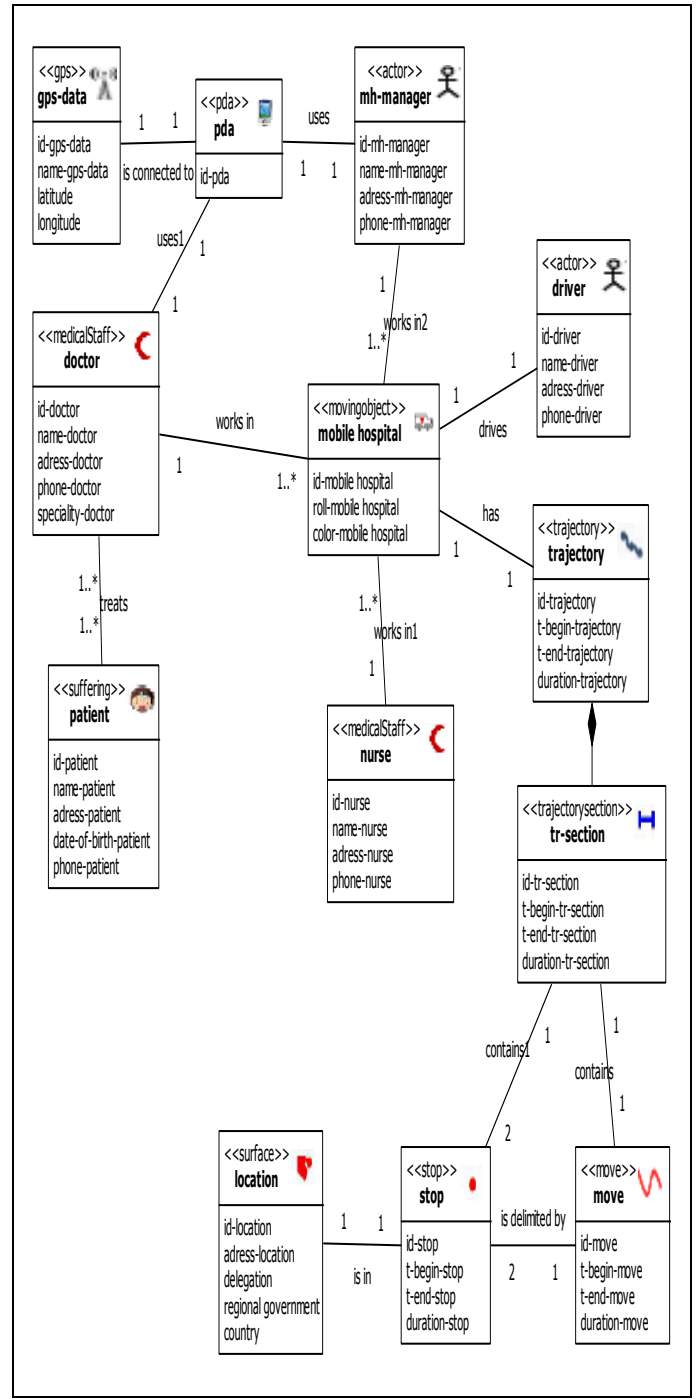

Figure 3. Mobile hospital's trajectory conceptual modeling

## IV. REALIZATION OF TRAJECTORY UML PROFILE

In our solution, we propose to add some extensions (stereotypes, icons) to UML diagrams (class diagram, sequence diagram). This leads to our trajectory profile which is based on UML. We were inspired by the work of [1, 15] to extend the UML profile in order to bring to light the trajectory and its environment.

An UML profile allows specializing UML in a precise domain. Our trajectory UML profile is composed of two diagrams which are the trajectory data sequence diagram and the trajectory data class diagram.

The first diagram has the aim to show how to consult a "patient" or to add some information to it. In the following table we describe each element of the trajectory data sequence diagram and the added stereotypes and icons.

Table I. Stereotypes and icons of the trajectory data sequence diagram

| Elements | Stereotypes | Icons |
|---|---|---|
| doctor | «MEDICALSTAFF» | |
| patient | «SUFFERING» | |
| UIpda | «userInterface» | |
| Control pda | «management» | |

The second diagram has the aim to show trajectory data classes and interactions between them. In the following table, we describe our new stereotypes and icons for each class which is in interaction with the trajectory class and the moving object class.

Table II. Stereotypes and icons of the trajectory data class diagram

| Elements | Stereotypes | Icons |
|---|---|---|
| Trajectory | «trajectory» | |
| Trajectory-section | «trajectory-section» | |
| Stop | «stop» | |
| Move | «move» | |
| Pda | «pda» | |
| Gps | «gps data» | |
| Location | «surface» | |
| Mobile hospital | «moving object» | |
| Doctor/nurse | « Medical staff » | |
| Driver/manager | «actor» | |
| Patient | «suffering» | |

To implement the trajectory data sequence diagram and the trajectory data class diagram, we used the open source platform called StarUml. This latter is extensible since it uses the XML. In fact, StarUML allows adding new functions which are adaptable to users' needs. We extended a new approach of UML called tdw (trajectory data warehouse) approach and a new profile. The "tdw approach" defines new types of diagrams (trajectory data sequence and class diagrams) and their order of appearance.

The trajectory UML profile is used to widen the capabilities of UML to express specific elements in a certain domain. The following schema shows the "tdw approach".

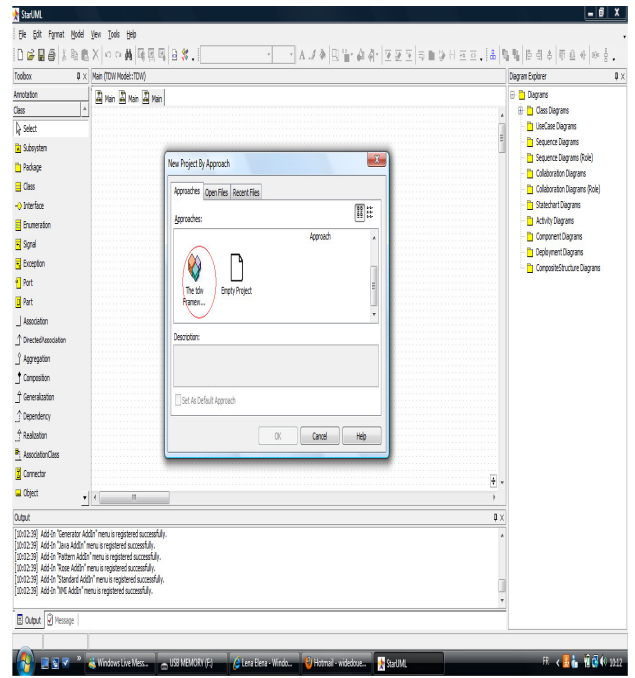

Figure 4. The tdw approach

After having chosen the "tdw approach", we find in the part "Model Explorer", the added model called TDW model which consists of two sub modes called trajectory data sequence diagram and trajectory data class diagram. The following schema shows the description cited above.

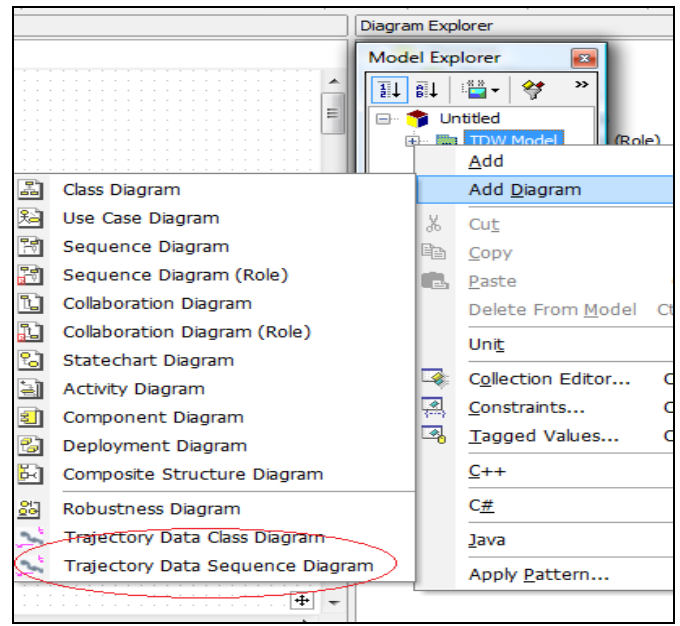

Figure 5. The trajectory data class and sequence diagrams

## V. CONCLUSION

In this work, we presented concepts related to trajectory data resulting from a specific mobile object which is the mobile hospital. We proposed a new profile called trajectory profile which is based on UML. In this trajectory profile, we defined a trajectory data sequence and a trajectory data class diagrams with new stereotypes and icons to concise the conceptual design of trajectory data. Such trajectory data resulting from mobile hospitals can be gathered and stored in a trajectory data warehouse then can be analyzed deeply using data mining tools to make strategic decision about some prospective epidemic peaks.